\documentclass{new_tlp}

\usepackage[semibold,scale=0.875]{sourcecodepro}
\usepackage[T1]{fontenc}
\usepackage[utf8]{inputenc}

\usepackage{amsmath}
\usepackage{amssymb}
\usepackage{enumerate}

\usepackage{hyperref}
\usepackage{microtype}
\usepackage{color}
\usepackage{listings}

\lstdefinelanguage{anthem}
{
	otherkeywords={\#},
	morekeywords={external, show, not},
	morecomment=[l]{\%},
	sensitive,
}

\lstdefinelanguage{FOL}
{
	morekeywords={and, or, not, exists, forall, in, int},
	morecomment=[l]{\%},
	sensitive,
}

\definecolor{gray}{rgb}{0.33,0.33,0.33}

\def\anthem{{\sc anthem}}
\def\gringo{{\sc gringo}}

\def\rar{\rightarrow}

\def\beq{\begin{equation}}
\def\eeq#1{\label{#1}\end{equation}}
\def\ba{\begin{array}}
\def\ea{\end{array}}

\title[anthem: Transforming gringo Programs into First-Order Theories]{anthem:
Transforming gringo Programs\\
into First-Order Theories (Preliminary Report)}

\author[Lifschitz, Lühne, and Schaub]
{
VLADIMIR LIFSCHITZ\\
University of Texas at Austin, USA\\
vl@cs.utexas.edu\\
\and
PATRICK LÜHNE\\
University of Potsdam, Germany\\
patrick.luehne@cs.uni-potsdam.de\\
\and
TORSTEN SCHAUB\\
University of Potsdam, Germany\\
INRIA Rennes, France\\
torsten@cs.uni-potsdam.de
}

\submitted{10 April 2018}
\revised{27 May 2018}
\accepted{27 May 2018}

\begin{document}
\label{firstpage}

\maketitle

\begin{abstract}
In a recent paper by Harrison et al., the concept of program completion
is extended to a large class of programs in the input language of the
ASP grounder \gringo{}. We would like to automate the process of generating
and simplifying completion formulas for programs in that language, because
examining the output produced by this kind of software may help programmers
to see more clearly what their program does and to what degree its set of
stable models
conforms with their intentions. If a formal specification for the program
is available, then it may be possible to use this software, in combination
with automated reasoning tools, to verify that the program is correct.
This note is a preliminary report on a project motivated by this idea.
 \end{abstract}

\section{Introduction}

Harrison et al.~\citeyear{har17a} extended the concept of program
completion~\cite{cla78} to a large class of nondisjunctive programs in the input
language of the ASP grounder \gringo{}~\cite{geb12}.  They argued that it
would be useful to automate the process of generating and simplifying
completion formulas for (tight\footnote{\emph{Tightness} is a syntactic condition
that guarantees the equivalence between the stable model semantics and the
completion semantics of a logic program~\cite{fag94,erd03}.})
\gringo{} programs, because
examining the output produced by this kind of software may help programmers
to see more clearly what their program does and to what degree its set of
stable models
conforms with their intentions.  Furthermore, if a formal specification
for a \gringo{} program is available, then it may be possible to use this
software, in combination with automated reasoning tools, to verify that the
program is correct.

This note is a preliminary report on a software development project that
follows up on this idea.  {\sc anthem} is a translator that converts a
\gringo{} program into its completion and simplifies it.  By \emph{simplifying} we
mean, in this case, not so much making formulas shorter as writing them in a
form that is ``readable''---natural from the perspective of a human who is
accustomed to expressing mathematical ideas using propositional connectives,
quantifiers, variables for objects of various types, the summation symbol,
and other standard notation.  The language of \gringo{} and many other input
languages of answer set solvers, including those of
{\sc smodels}~\cite{nie97} and {\sc dlv}~\cite{leo06a}, classify
variables into global and local, instead of using quantifiers to classify
occurrences into free and bound, and that distinguishes them from
traditional notation.  The same can be said about assuming that all
variables range over the same universe, instead of using variables of
different sorts or types (for points, lines, and planes; for integers and
real numbers; or for sets and classes; etc.).\footnote{Answer set
programming with \emph{sorts}~\cite{bal13} is
an exception.  What these authors proposed, however, is assigning sorts to
argument positions but not assigning them to variables.  This is different
from what
is customary in logic and in procedural programming languages.}  Each of
the two notational traditions has its advantages, and \anthem{} provides a
bridge between them.

Simplifying a logical formula in the sense of making it more understandable
is not the same thing, of course, as rewriting it using fewer characters.  For
instance, the equivalent formulas
$$\exists x\forall y(P(y) \rar Q(x))$$
and
$$\exists xP(x) \rar \exists xQ(x)$$
use the same number of logical symbols, but the latter is much
easier to understand.

Besides generating and simplifying the completion of a program, \anthem{}
``hides'' auxiliary predicate symbols occurring in the program when
possible.  In the language of \gringo, the fact that a predicate symbol is
not considered an essential part of the output can be expressed
by not including it in \texttt{\textbf{\#show}} directives.  To eliminate such
predicate symbols from its output, \anthem{} replaces them by their
completed definitions.

The input language of \anthem{} is a large part of the input
language of \gringo.  Input programs are supposed to be
nondisjunctive.  They may use arithmetic operations, intervals,
comparisons, singleton choice rules without lower and upper bounds, and
constraints.  Aggregates and conditional
literals are not supported in the current version.

The output of {\sc anthem} is a list of first-order formulas with variables
of two sorts---for arbitrary precomputed terms (that is, for all elements of
the Herbrand universe) and for the precomputed terms that correspond to
integers---as proposed by Harrison et al.~\citeyear[Sections~3 and 9]{har17a}.
Differences between atoms in \gringo{} programs and atomic parts of
formulas are related mostly to arithmetic expressions (see
Section~\ref{sec:syntax} below).

The GitHub repository of \anthem{}
contains the source code, installation steps, usage instructions, as well as
multiple examples to experiment with.\footnote{\url{https://github.com/potassco/anthem}}

\pagebreak
\section{Examples} \label{sec:examples1}

\noindent{\it Example 1.}
Given the input file
\begin{lstlisting}[language=anthem]
  s(X) :- p(X).
  s(X) :- q(X).

  #external p(1).
  #external q(1).
\end{lstlisting}
\anthem{} generates the formula
\begin{lstlisting}[language=FOL]
  forall V1 (s(V1) <-> (p(V1) or q(V1)))
\end{lstlisting}
The first two lines of the input file express the condition
$$s=p\cup q$$
in the language of logic programming.  The last two lines tell us that
the unary predicates~{\tt p} and~{\tt q} are not defined in this file;
they can be specified separately.\footnote{This is slight abuse of the
syntax of \texttt{\textbf{\#external}} directives, since a predicate's arity is
indicated by its argument.}  (Without those lines in the input, \anthem{}
would have decided that {\tt p} and {\tt q} are identically false.) The
output of \anthem, shown above, expresses the same condition by a
first-order formula.

\medskip\noindent{\it Example 2.}
The condition
$$u=r\setminus(p\cup q)$$
can be represented by the \gringo{} program
\begin{lstlisting}[language=anthem]
  s(X) :- p(X).
  s(X) :- q(X).
  u(X) :- r(X), not s(X).

  #show u/1.

  #external p(1).
  #external q(1).
  #external r(1).
\end{lstlisting}
\anthem{} converts this program into the formula
\begin{lstlisting}[language=FOL]
  forall V1 (u(V1) <-> (r(V1) and not (p(V1) or q(V1))))
\end{lstlisting}
which expresses the same condition in traditional logical notation.  It is
obtained from the completed definition
\begin{lstlisting}[language=FOL]
  forall V1 (u(V1) <-> (r(V1) and not s(V1)))
\end{lstlisting}
of {\tt u} by replacing {\tt s} with its completed definition.

\medskip\noindent{\it Example 3.}  The rules
\begin{lstlisting}[language=anthem]
  p(a).  {q(a)}.
\end{lstlisting}
express---assuming that {\tt p} and {\tt q} do not occur in the heads of
any other rules---that $p=\{a\}$ and $q\subseteq\{a\}$.  The formulas
\begin{lstlisting}[language=FOL]
  forall V1 (p(V1) <-> V1 = a)
  forall V2 (q(V2) -> V2 = a)
\end{lstlisting}
generated by \anthem{} in response to this input express the same conditions
in traditional notation.

\medskip\noindent{\it Example 4.}  The vertex coloring program
\begin{lstlisting}[language=anthem]
  % assign a set of colors to each vertex
  {color(V, C)} :- vertex(V), color(C).
  
  % at most one color per vertex
  :- color(V, C1), color(V, C2), C1 != C2.
  
  % at least one color per vertex
  colored(V) :- color(V, _).
  :- vertex(V), not colored(V).
  
  % adjacent vertices don't share the same color
  :- color(V1, C), color(V2, C), edge(V1, V2).
  
  #show color/2.
  
  #external vertex(1).
  #external edge(2).
  #external color(1).
\end{lstlisting}
is transformed by \anthem{} into
\begin{lstlisting}[language=FOL]
  forall V1, V2 (color(V1, V2) -> (vertex(V1) and color(V2)))
  forall U1, U2, U3 (not color(U1, U2) or not color(U1, U3) or U2 = U3)
  forall U4 (vertex(U4) -> exists U5 color(U4, U5))
  forall U6, U7, U8 (not color(U6, U7) or not color(U8, U7)
    or not edge(U6, U8))
\end{lstlisting}
The second of these formulas will look more natural if we rewrite it as
\begin{lstlisting}[language=FOL]
  forall U1, U2, U3 ((color(U1, U2) and color(U1, U3)) -> U2 = U3)
\end{lstlisting}
and the last of them can be improved by writing it in the form
\begin{lstlisting}[language=FOL]
  not exists U6, U7, U8 (color(U6, U7) and color(U8, U7)
    and edge(U6, U8)).
\end{lstlisting}
These simplifications will be implemented in a future version of \anthem{}.

\pagebreak
\section{Arithmetic Expressions in Formulas}\label{sec:syntax}

In the output of \anthem{}, an integer variable can be recognized by its
first character---the letter {\tt N}.  For instance, the formula
$$\texttt{\textbf{exists} N p(N)}$$
is stronger than
$$\texttt{\textbf{exists} X p(X)}$$
---it expresses that the set {\tt p} contains at least one integer (and not
only ground terms formed using symbolic constants).

In the language of \gringo{}, a ground term represents, generally, a finite
set of values, rather than a single value~\cite[Section~A.1]{har17a}.
For example, the term \texttt{1}~\texttt{+}~\texttt{3} has a single
value {\tt 4}, but \texttt{1..3} has the values {\tt 1}, {\tt 2}, {\tt 3}. The
set of values of \texttt{3..1} is empty, and so is the set of values of
\texttt{3}~\texttt{/}~\texttt{(1}~\texttt{-}~\texttt{1)}.
Accordingly, an atomic formula can be formed from two terms using the
symbol \texttt{in}, expressing set membership, for instance:
$$
\texttt{X \textbf{in} 1 + 3},\quad\texttt{X \textbf{in} 1..3},\quad\texttt{X \textbf{in} 3..1},\quad
\texttt{X \textbf{in} 3 / (1 - 1)}.
$$
The first of these formulas is equivalent to \texttt{X}~\texttt{=}~\texttt{4}; the second, to
$$\texttt{X = 1 \textbf{or} X = 2 \textbf{or} X = 3};$$
the last two, to \texttt{\textbf{\#false}}.

On the other hand, intervals are allowed in a formula in only one
position---to
the right of the symbol \texttt{in}.   For instance, the atom \texttt{p(1..3)},
which can be used in \gringo{} rules,
is not a formula; in the world of formulas, we distinguish between
$$\texttt{\textbf{exists} X (X \textbf{in} 1..3 \textbf{and} p(X))}$$
and
$$\texttt{\textbf{forall} X (X \textbf{in} 1..3 -> p(X))}.$$
The use of arithmetic operations in formulas is restricted in the same way,
except for terms formed from integer variables using addition, subtraction,
and multiplication.  Such terms can be used in a formula anywhere.  This
exception is motivated by the fact that any ground term of this type has a
unique value.  For
instance, \texttt{p(N}~\texttt{+}~\texttt{1)} is a formula, but
\texttt{p(X}~\texttt{+}~\texttt{1)} is not.  To express that
$$\hbox{the value of {\tt X} is an integer, and its successor belongs
to {\tt p}},$$
we write
$$\texttt{\textbf{exists} N (X = N \textbf{and} p(N + 1))}.$$

In the output of \anthem, an expression of the form \texttt{\textbf{int}(p/n@k)}, where $p$ is a
symbolic constant and $n$, $k$ are integers such that $1\leq k\leq n$, stands
for the formula
$$
\texttt{\textbf{forall}}~X_1,\dots, X_n~(p(X_1,\dots, X_n)~\texttt{->}~\texttt{\textbf{exists}}~N~(X_k = N)),
$$
which expresses that the $k$-th member of any $n$-tuple satisfying $p$ is an
integer.

\pagebreak
\section{Examples Involving Arithmetic Expressions}\label{sec:examples2}

\noindent{\it Example 5.}
The program
\begin{lstlisting}[language=anthem]
  letter(a).  letter(b).  letter(c).
  {p(1..3, Y)} :- letter(Y).
  :- p(X1, Y), p(X2, Y), X1 != X2.
  q(X) :- p(X, _).
  :- X = 1..3, not q(X).

  #show p/2.
\end{lstlisting}
encodes the set of permutations of the letters $a$, $b$, $c$.  \anthem{}
transforms it into
\begin{lstlisting}[language=FOL]
  forall N1, V1 (p(N1, V1) -> (N1 in 1..3 and (V1 = a or V1 = b
    or V1 = c)))
  forall N2, U1, N3 (not p(N2, U1) or not p(N3, U1) or N2 = N3)
  forall N4 (N4 in 1..3 -> exists U2 p(N4, U2))
  int(p/2@1)
\end{lstlisting}

\medskip\noindent{\it Example 6.}
The program
\begin{lstlisting}[language=anthem]
  composite(I * J) :- I = 2..n, J = 2..n.
  prime(N) :- N = 2..n, not composite(N).

  #show prime/1.
\end{lstlisting}
encodes the set of primes up to $n$. \anthem{} turns it into
\begin{lstlisting}[language=FOL]
  forall N1 (prime(N1) <-> (N1 in 2..n and not exists N2, N3
    (N1 = (N2 * N3) and N2 in 2..n and N3 in 2..n)))
  int(prime/1@1)
\end{lstlisting}

\medskip\noindent{\it Example 7.}
The program
\begin{lstlisting}[language=anthem]
  {in(1..n, 1..r)}.
  covered(I) :- in(I, _).
  :- I = 1..n, not covered(I).
  :- in(I, S), in(J, S), in(I + J, S).

  #show in/2.
\end{lstlisting}
encodes partitions of $\{1,\dots, n\}$ into~$r$ sum-free sets. \anthem{} turns
it into
\begin{lstlisting}[language=FOL, escapeinside=\{\}]
  forall N1, N2 ({in}(N1, N2) -> (N1 in 1..n and N2 in 1..r))
  forall N3 (N3 in 1..n -> exists N4 {in}(N3, N4))
  forall N5, N6, N7 (not {in}(N5, N6) or not {in}(N7, N6)
    or not {in}((N5 + N7), N6))
  int({in}/2@1)
  int({in}/2@2)
\end{lstlisting}

\section{Implementation}
\label{sec:system}

The implementation of \anthem{} takes advantage of \gringo's library
functionality for accessing the abstract syntax tree (AST) of a nonground
program.  The AST obtained from \gringo{} is taken by \anthem{} and
turned into the AST of the collection of formulas representing the rules
of the program~\cite[Section~4]{har17a}.  That tree is then turned into the
AST of the program's completion.  Auxiliary predicates are eliminated, and
the result is subject to several simplifications, including those that
involve the use of integer variables.
In Example~5, for instance, the formula
\begin{lstlisting}[language=FOL]
  forall V2, V3 (p(V2, V3) -> (V2 in 1..3 and letter(V3)))
\end{lstlisting}
is replaced at the last stage by the pair of formulas
\begin{lstlisting}[language=FOL]
  forall N1, V2 (p(N1, V2) -> (N1 in 1..3 and letter(V2))),
  int(p/2@1).
\end{lstlisting}
The final result is pretty-printed to standard output.

In the process of eliminating auxiliary predicates by replacing them with
their definitions, a cycle check is employed to detect the cases when this
process would not terminate.  For example, in response to the input
\begin{lstlisting}[language=anthem]
  p :- not q.
  q :- not p.
  r :- not p.

  #show r/0.
\end{lstlisting}
\anthem{} displays a warning:
$$\verb|cannot hide predicate “q/0” due to circular dependency|.$$

Apart from the standard options (for help, version, etc.),
\anthem{} allows us to switch on/off completion, simplification, and
introducing integer variables from the command line.

\section{Future Work}

Future work on \anthem{} will proceed in two main directions.  First, we
would like to support aggregates and conditional literals.  When this is
accomplished, we will be able to replace, for instance, the first four rules
of Example~4 by a single rule
\begin{lstlisting}[language=anthem]
  1 {color(V, C) : color(C)} 1 :- vertex(V).
\end{lstlisting}
With aggregates added to the input language of \anthem{}, formulas in
its output will have more complex syntactic
structure~\cite[Section~8.2]{har17a}.

Second, we will investigate the possibility of using automated
reasoning tools for classical logic, in combination with \anthem, for
verifying programs written in the input language of \gringo.  For
instance, the specification for the program from Example~6---encode the set
of primes up to $n$---can be expressed by the formula
\begin{lstlisting}[language=FOL]
  forall N1 (prime(N1) <-> (1 < N1 <= n and not exists N2, N3
    (N1 = N2 * N3 and N2 > 1 and N3 > 1))).
\end{lstlisting}
It is easy to show that this formula is equivalent to the first formula
in the output of anthem using first-order logic, simple properties of
inequalities, and the meaning of the interval notation.  We expect that
it will be possible to verify claims of this kind using a proof checker.

\section*{Acknowledgements}

Thanks to Amelia Harrison and to the anonymous referees for comments
on a draft of this note.
The first author was partially supported by the National Science Foundation
under Grant IIS-1422455.

\bibliographystyle{acmtrans}

\end{document}